\documentclass[manuscript]{aastex}

\usepackage{rotating}
\usepackage[dvips]{color}
\usepackage{verbatim}
\usepackage{soul}
\usepackage{longtable}
\usepackage{subfigure}
\usepackage{amsmath}

\slugcomment{To be submitted to AJ Main Journal. Version}

\shorttitle{Beyond Frequency Phase Transfer}

\shortauthors{G.-Y. Zhao, J. C. Algaba and the iMOGABA collaboration}

\begin{document}

\title{The Power of Simultaneous Multi-frequency Observations \\
for mm-VLBI: Beyond Frequency Phase Transfer} 

\author{
Guang-Yao~Zhao$^{1}$,
Juan~Carlos Algaba$^{1}$,
Sang~Sung~Lee$^{1,2}$,
Taehyun~Jung$^{1,2}$
Richard~Dodson$^{3}$,
Mar\'ia Rioja$^{3,4,5}$,
Do-Young Byun$^{1,2}$,
Jeffrey Hodgson$^{1}$,
Sincheol Kang$^{1,2}$,
Dae-Won Kim$^{6}$,
Jae-Young Kim$^{7}$,
Jeong-Sook Kim$^{8}$,
Soon-Wook Kim$^{1,2}$,
Motoki Kino$^{1,8,9}$,
Atsushi Miyazaki$^{1,10}$,
Jong-Ho Park$^{6}$,
Sascha Trippe$^{6}$,
Kiyoaki Wajima$^{1}$}
\affil{
$^1$Korea Astronomy \& Space Science Institute, 776, Daedeokdae-ro, Yuseong-gu, Daejeon 34055, Korea\\
$^2$University of Science and Technology, 217, Gajeong-ro, Yuseong-gu, Daejeon 34113 , Korea\\
$^3$ICRAR, M468, University of Western Australia, 35 Stirling  Hwy, Perth 6009, Australia\\
$^4$CSIRO Astronomy and Space Science, 26 Dick Perry Avenue, Kensington WA 6151, Australia\\
$^5$OAN (IGN), Alfonso XII, 3 y 5, 28014 Madrid, Spain\\
$^6$Department of Physics and Astronomy, Seoul National University, Gwanak-ro, Gwanak-gu, Seoul 08826, Korea\\
$^7$Max-Planck-Institut f\"ur Radioastronomie, Auf dem H\"ugel 69, 53121 Bonn, Germany\\
$^8$National Astronomical Observatory of Japan, 2-21-1 Osawa, Mitaka, Tokyo 181-8588, Japan\\
$^9$Kogakuin University, Academic Support Center, 2665-1 Nakano, Hachioji, Tokyo 192-0015, Japan\\
$^{10}$Faculty of Science and Engineering, Hosei University, 3-7-2 Kajino-cho, Koganei, Tokyo 184-8584, Japan\\
}

\begin{abstract}

Atmospheric propagation effects at millimeter wavelengths can significantly alter the phases of radio signals and reduce the coherence time, 
 putting tight constraints on high frequency Very Long Baseline Interferometry (VLBI) observations. 
In previous works, it has been shown that non-dispersive (e.g. tropospheric) effects can be calibrated with the frequency phase transfer (FPT) technique.
The coherence time can thus be significantly extended. 
Ionospheric effects, which can still be significant, remain however uncalibrated after FPT as well as the instrumental effects. In this work, we implement a further phase transfer between two FPT residuals (i.e.~{\it so-called} FPT--square) to calibrate the ionospheric effects based on their frequency dependence. 
We show that after FPT--square, the coherence time at 3 mm can be further extended beyond 8~hours, and the residual phase errors can be sufficiently canceled by applying the calibration of another source, which can have a large angular separation from the target ($>20\degr$) and significant temporal gaps. 
Calibrations for all--sky distributed sources with a few calibrators are also possible after FPT--square.
One of the strengths and uniqueness of this calibration strategy is the suitability for high frequency all-sky survey observations including very weak sources. 
We discuss the introduction of a pulse calibration system in the future to calibrate the remaining instrumental effects and allowing the possibility of imaging the source structure at high frequencies with FPT--square, where all phases are fully calibrated without involving any additional sources.

\end{abstract}


\section{Introduction}

Very long baseline interferometry (VLBI) techniques have allowed astronomical observations of active galactic nuclei (AGNs) to reach the highest resolutions so far, down to few tens of microarcseconds \citep[see e.g.][]{Doeleman08, Gomez16}. However, although this technique is very well established at centimeter wavelengths, it becomes very challenging when trying to achieve better resolutions at shorter (millimeter and sub-millimeter) wavelengths. This is not only due to the typically lower fluxes of synchrotron radiation at shorter wavelengths, but also because of a combination of observational effects, from the antenna optics and receiver system to the atmospheric transparency and stability.

Indeed, at higher frequencies (short mm-wavelengths), the impact of the antenna surface accuracy is more relevant to the observations and systems with a relatively low receiver temperature are hard to implement. This leads to higher effective system temperatures and system equivalent flux densities (SEFDs), and irredeemably, poorer sensitivities. In addition, the signal is degraded in different ways as it passes through the atmosphere: attenuation may occur due to atmospheric absorption, scattering and refraction occur due to variations in the water vapour distribution in the troposphere that moves across the interferometer. In particular, the latest effect prevents long coherence times and effective integration times of no more than a few minutes.
As a result, the number of detectable sources has been very limited at radio frequencies higher than 86 GHz~\citep[see e.g.][]{Lee08}.

Despite these difficulties, there is an obvious interest in VLBI observations at such high frequencies (e.g. 86~GHz or higher). As most sources become optically-thin \citep[e.g.][]{iMOGABA-I_paper}, it is much easier to study their central regions at these frequencies. Such observations can probe upstream of the jets and into the vicinities of the AGN central engines, thus revealing key information about their source of energy, jet footpoint conditions, acceleration and collimation~\citep[e.g.][]{hada16,hodgson17}.

Efforts towards (sub)millimeter--VLBI observations have thus been done. Three of the most noteable examples are the global millimeter VLBI array (GMVA)\footnote{http://www3.mpifr-bonn.mpg.de/div/vlbi/globalmm/}, which can observe to up to 86~GHz, the Korean VLBI network (KVN)\footnote{http://radio.kasi.re.kr/kvn/main\_kvn.php}, with observations up to 129~GHz, and the Event Horizon Telescope (EHT)\footnote{http://www.eventhorizontelescope.org/} which has observed at 230~GHz. High frequency VLBI will greatly benefit from the inclusion of the phased Atacama Large Millimeter Array (ALMA). Note that such systems are in the current technological frontier.



In order to increase the efficiency of high frequency observations, atmospheric phase corrections are needed. For this, the use of additional techniques such as water vapor radiometry (WVR) \citep{Stirling2006} are required. For example, the 183~GHz water vapor radiometer installed at an ALMA 12~m antenna has been extremely successful \citep[e.g.,][]{matsushita17}. On the other hand, for many facilities, WVR may not be available and other (supplementary) techniques are welcome. 

In this sense, the KVN is a very special facility. With its unique quasi--optics system designed to split one signal from the sub--reflector into different frequency bands and guide them to the corresponding receivers, up to four frequencies can be observed simultaneously \citep{Han08,Han13}.
One of the techniques that can take great advantage of simultaneous observations is the frequency phase transfer (FPT). The non--dispersive tropospheric fluctuations of the atmosphere at higher frequencies are corrected by transferring phase solutions from a lower frequency which is observed simultaneously \citep[e.g.][]{Asaki98, perez2010, Jung11PASJ} or close in time~\citep[e.g.][]{Middelberg05, Dodson09}. Simultaneous observing can be a great advantage in, among other things, preventing time interpolation errors~\citep[e.g.][]{RD2011, RD2014}. The FPT technique has been broadly discussed in e.g. \cite{RD2011} and is now extensively used in VLBI observations up to 129~GHz \citep[e.g.][]{RD2015, Algaba15}. FPT for KVN 4-frequency continuum observation data can now be performed automatically with the KVN pipeline ~\citep{kvnpipeline}. 

We note however that the FPT technique has its own limitations. The most obvious limitation is that it does not correct for the dispersive effects (e.g. ionospheric, which may still be significant once other effects have been accounted for). Moreover, these effects at the reference frequency are scaled up and brought into the target frequencies. In practice this means that coherence times will still be limited to a few tens of minutes, and there may still be a clear line of sight dependence of the phases. 

To overcome this limitation, the most direct way is to use self--calibration with the extended coherence time. However, even after FPT, there are still weak sources that cannot be reliably detected at high frequencies (86, and 129 GHz) with self--calibration \citep{Algaba15}.
Another method is to use an additional nearby bright source as a phase calibrator, known as source frequency phase referencing~\citep[SFPR;][]{RD2011}. In this case the lines of sight should be be close enough to guarantee a reliable calibration. Alternatively, \cite{Dodson17} developed an advanced multi--frequency phase--referencing (MFPR) in which ionospheric execution blocks are included to obtain the ling-of-sight Total Electron Content (TEC) and the need of a nearby phase calibrator is effectively removed.

    In this paper, we introduce a new approach to circumvent these limitations. We extend the applicability of the FPT technique in order to take into account the ionospheric effects, increasing the coherence time to periods longer than a few hours, and perform a further phase residual correction on all--sky sources for the remaining instrumental effects. The contents are distributed as follows: Section 2 describes the methodology, Section 3 summarizes the observations and data analysis, Section 4 presents the results, which are discussed in Section 5, and Section 6 summarizes the conclusions.

\section{Methods}

In order to calibrate atmospheric propagation errors, several techniques have been used in the past~\citep[e.g.][]{BeasleyConway95, Asaki98}. Among them, FPT has provided striking results by utilizing simultaneous multi--frequency data~\citep[e.g.][]{RD2011, RD2015, Algaba15}. However, whereas this technique can successfully calibrate tropospheric effects, other effects, such as ionospheric and instrumental, are still present. Although ionospheric effects may not be important compared with tropospheric ones at mm-wavelengths, they may still dominate over instrumental ones, depending on the array performance. 

Nonetheless, the same basic principles of the FPT technique can be extended for the correction of ionospheric effects given that they also follow a well--defined dependence with frequency. Ionospheric phase effects are inversely proportional to frequencies, as opposed to tropospheric ones. 

In this section we describe the basics of the FPT technique for correcting the tropospheric effects and its extension into the {\it so-called} FPT--square which also takes into account the ionospheric effects. In the following we will only consider facilities with simultaneous multi--frequency capabilities such as KVN for simplicity. 

\subsection{VLBI Phase Errors}
Following standard nomenclature, the visibility phase of a VLBI observation of a source A at each frequency consists of various terms which can be described as contributions arising from the source structure, atmosphere, and instruments, among others:

\begin{eqnarray}
  \phi^{{\rm \nu_{1}}}_{{\rm A}} = \phi^{{\rm \nu_{1}}}_{{\rm A,str}} + \phi^{{\rm \nu_{1}}}_{{\rm A,geo}} + 
  \phi^{{\rm \nu_{1}}}_{{\rm A,tro}} +\phi^{{\rm \nu_{1}}}_{{\rm A,ion}}
  +\phi^{{\rm \nu_{1}}}_{{\rm A,inst}} + \phi^{{\rm \nu_{1}}}_{{\rm A,thermal}} + 
  2\pi n^{{\rm \nu_{1}}}_{{\rm A}}, \nonumber \\
  \phi^{{\rm \nu_{2}}}_{{\rm A}} = \phi^{{\rm \nu_{2}}}_{{\rm A,str}} + \phi^{{\rm \nu_{2}}}_{{\rm A,geo}} + 
  \phi^{{\rm \nu_{2}}}_{{\rm A,tro}} +\phi^{{\rm \nu_{2}}}_{{\rm A,ion}}
  +\phi^{{\rm \nu_{2}}}_{{\rm A,inst}} +  \phi^{{\rm \nu_{2}}}_{{\rm A,thermal}} +  2\pi n^{{\rm \nu_{2}}}_{{\rm A}},\\
  \phi^{{\rm \nu_{3}}}_{{\rm A}} = \phi^{{\rm \nu_{3}}}_{{\rm A,str}} + \phi^{{\rm \nu_{3}}}_{{\rm A,geo}} + 
  \phi^{{\rm \nu_{3}}}_{{\rm A,tro}} +\phi^{{\rm \nu_{3}}}_{{\rm A,ion}}
  +\phi^{{\rm \nu_{3}}}_{{\rm A,inst}} +  \phi^{{\rm \nu_{3}}}_{{\rm A,thermal}} +  2\pi n^{{\rm \nu_{3}}}_{{\rm A}}, \nonumber
\end{eqnarray} 

where $\phi^{{\rm \nu_{1}}}_{{\rm A}}$, $\phi^{{\rm \nu_{2}}}_{{\rm A}}$ and $\phi^{{\rm \nu_{3}}}_{{\rm A}}$ stand for the observed visibility phases at three different frequencies; $\phi_{{\rm A,str}}$ arises from the source structure; $\phi_{{\rm A,geo}}$, $\phi_{{\rm A,tro}}$, $\phi_{{\rm A,ion}}$, $\phi_{{\rm A,inst}}$ are the contributions arising from geometric, tropospheric, ionospheric, and instrumental residuals of the model used by the correlator, respectively; $\phi_{{\rm A,thermal}}$ stands for thermal noise errors related with array sensitivity, and the last term represents the $2\pi$ ambiguity of the phase with $n_{{\rm A}}^{{\nu_i}}$ an integer value at each frequency.

\subsection{First-order Frequency Phase Transfer (FPT)}
As the step, we apply the standard FPT to achieve the calibrations of non-dispersive phase effects at $\nu_2$ and $\nu_3$ based on the solutions found at $\nu_1$. Details of this procedure are  well described in \citet{RD2011}. Here, we briefly describe the basics: due to its non-dispersive nature, the tropospheric term is linearly proportional to the observing frequency, and thus it is possible to calibrate it at one frequency with simultaneously obtained data at a different frequency. The finge--fitting phase solutions $\phi^{{\rm \nu_{1}}}_{{\rm A,fring}}$ contain all the contributing terms except the source structure. By multiplying the frequency ratio, $R_{21} = \nu_{2}/\nu_{1}$ and eliminating it from $\phi_{\rm A}^{{\rm \nu_{2}}}$, the residual phase becomes

\begin{center}
\begin{eqnarray}
\label{residualPhaseEq}
\phi_{\rm A,FPT}^{{\rm \nu_{2}}} = 
\phi_{\rm A}^{{\rm \nu_{2}}} - R_{21}\,.\,\phi_{{\rm A,fring}}^{{\rm \nu_{1}}} 
= \phi_{{\rm A,str}}^{{\rm \nu_{2}}} + (\phi_{{\rm A,geo}}^{{\rm \nu_{2}}} - R_{21}\,.\,\phi_{{\rm A,geo}}^{{\rm \nu_{1}}}) 
+ (\phi_{{\rm A,tro}}^{{\rm \nu_{2}}} - R_{21}\,.\,\phi_{{\rm A,tro}}^{{\rm \nu_{1}}}) \nonumber \\
+ (\phi_{{\rm A,ion}}^{{\rm \nu_{2}}} - R_{21}\,.\,\phi_{{\rm A,ion}}^{{\rm \nu_{1}}})  
+ (\phi_{{\rm A,inst}}^{{\rm \nu_{2}}} - R_{21}\,.\,\phi_{{\rm A,inst}}^{{\rm \nu_{1}}}) 
+ 2\pi (n^{{\rm \nu_{2}}}_{\rm A}-R_{21}\,.\,n^{{\rm \nu_{1}}}_{\rm A})  ,
\end{eqnarray} 
\end{center}



where we have ignored the thermal noise term for simplicity as it is usually small for typical KVN observations because i) the sources should be bright at the reference frequency (e.g. brighter than 180 mJy at 22 GHz\footnote{Assuming a recoding bandwidth of 64~MHz and integration time of 30 seconds, these give a signal-to-noise ratio higher than 5 under typical weather conditions}) to ensure a successful FPT, ii) at later steps, the thermal noise can be  effectively reduced by integrating over longer time scales because the coherence time is extended to several tens of minutes by FPT~\citep[e.g.][]{RD2015, Algaba15} and can be further extended up to hours by SFPR~\citep{RD2015} and the proposed method in this work (see below). 

After this step, the tropospheric term will be cancelled out, as well as other non-dispersive terms in the geometric residual phase (e.g. those arises from inaccurate antenna and source coordinates). The residual geometric phase comes from the position difference between the two frequencies $2\pi\, \vec{D}_{\lambda_{{\rm 2}}}\,.\, \vec{\theta}^{{\rm \nu_{1} \nu_{2}}}_{{\rm A}}$, where $2\pi\, \vec{D}_{\lambda_{{\rm 2}}}$ and $\vec{\theta}^{{\rm \nu_{1} \nu_{2}}}_{{\rm A}}$ are the interferometric baseline vector in units of $\lambda_2$ and the target source core shift  between the two frequencies, respectively~\citep{Lobanov98}.  Equation \ref{residualPhaseEq} becomes:

\begin{equation}
\label{eq:FPT12}
\phi_{\rm A,FPT}^{{\rm \nu_{2}}} = 
 \phi_{{\rm A,str}}^{{\rm \nu_{2}}} + 
2 \pi \, \vec{D}_{\lambda_{{\rm 2}}}\,.\, \vec{\theta}^{{\rm \nu_{1} \nu_{2}}}_{{\rm A}}
+ (\frac{1}{R_{21}}-R_{21} ) \, \phi_{{\rm A,ion}}^{{\rm \nu_{1}}} 
+ (\phi_{{\rm A,inst}}^{{\rm \nu_{2}}} - R_{21} \,.\, \phi_{{\rm A,inst}}^{{\rm \nu_{1}}}),
\end{equation}

Similarly, the residual phase at $\nu_{3}$ after FPT from $\nu_{1}$ is
\begin{equation}
\label{eq:FPT13}
\phi_{\rm A,FPT}^{{\rm \nu_{3}}} = 
 \phi_{{\rm A,str}}^{{\rm \nu_{3}}} + 
2 \pi \, \vec{D}_{\lambda_{{\rm 3}}}\,.\, \vec{\theta}^{{\rm \nu_{1} \nu_{3}}}_{{\rm A}}
+ (\frac{1}{R_{31}}-R_{31} ) \, \phi_{{\rm A,ion}}^{{\rm \nu_{1}}} 
+ (\phi_{{\rm A,inst}}^{{\rm \nu_{3}}} - R_{31} \,.\, \phi_{{\rm A,inst}}^{{\rm \nu_{1}}}),
\end{equation}

FPT is very powerful in expanding the coherence time at mm-wavelengths as mentioned above. However, as shown in Equations \ref{eq:FPT12} and \ref{eq:FPT13}, the dispersive (i.e. ionospheric and instrumental) terms remain in the phases which prevent imaging of the source at this step, even though the ionospheric effects also have frequency dependence ($\phi_{{\rm ion}} \propto \nu^{-1}$).

\subsection{Second-order Frequency Phase Transfer (FPT--square)}
As the second step, several methods have been suggested by previous works such as self-calibrating with the longer coherence times \citep[e.g.][]{Jung11PASJ, Algaba15} or calibrating using another source (i.e., SFPR) which additionally provides astrometry \citep[e.g.][]{RD2011, RD2015}. Here we propose another approach to calibrate the ionospheric effects. By applying an additional FPT between the residuals of the already FPT-ed phases (i.e. between Equations \ref{eq:FPT12} and \ref{eq:FPT13}), we get

\begin{eqnarray} \label{eq:FPTsqr}
\begin{aligned}
   \phi_{\rm A,FPT^{2}}^{{\rm \nu_{3}}} & = 
\phi_{\rm A,FPT}^{{\rm \nu_{3}}} - R\,.\,\phi_{{\rm A,FPT}}^{{\rm \nu_{2},fring}} \\
  & = \phi_{{\rm A,str}}^{{\rm \nu_{3}}} + 
2 \pi \, (\vec{D}_{\lambda_{{\rm 3}}}\,.\, \vec{\theta}^{{\rm \nu_{1} \nu_{3}}}_{{\rm A}} - R \,. \, \vec{D}_{\lambda_{{\rm 2}}}\,.\, \vec{\theta}^{{\rm \nu_{1} \nu_{2}}}_{{\rm A}}) \\
  & + [(\phi_{{\rm A,inst}}^{{\rm \nu_{3}}} - R_{31} \,.\, \phi_{{\rm A,inst}}^{{\rm \nu_{1}}}) - R\,.\, (\phi_{{\rm A,inst}}^{{\rm \nu_{2}}} - R_{21} \,.\, \phi_{{\rm A,inst}}^{{\rm \nu_{1}}})] 
\end{aligned}
\end{eqnarray}

with $R = (\frac{1}{R_{31}} - R_{31})/(\frac{1}{R_{21}} - R_{21})$. Hereafter, we will refer to this step as the FPT--square. One can easily see from Equation \ref{eq:FPTsqr} that the ionospheric (and in general any dispersive effect with $\nu^{-1}$ functional dependence) part cancels out due to its frequency dependence and only the instrumental terms are left uncalibrated, together with other quantities of interest such as source structure and geometric information which is a weighted combination of the core-shifts at two frequency pairs.

Ideally, the instrumental effects are not difficult to remove. A prime calibration using a short scan of a bright source could remove the instrumental offset at the time of this scan at different frequency bands~\citep{manualpcal}, and the residual instrumental effects do not have strong variations with time or source position. This will result in i) a very long coherence time, even at high frequencies (e.g. 86~GHz) and ii) feasible calibration with another source(s) even with a large separation between calibrator and target sources ($\sim20\degr$ or more, see Section~\ref{sec:results}), after FPT--square calibration is performed. The weak source detection at such frequencies will greatly benefit from this long coherence time and large allowed separation from calibrators. 

Furthermore, the residual instrumental variations can be monitored by the implementation of a pulse calibration (P-cal) system. Such system measures the phase and group delay in the receiver by generating short impulses in front of the feed~\citep[e.g.][]{Thompson91}. A multi-frequency P-Cal system will further improve the instrumental stability and broaden the scope of application of FPT--square. 

\section{Observations and Data Analysis}

In order to test the feasibility of FPT--square we searched in the archival for a suitable observation. We chose the 16th epoch of the KVN interferometric monitoring of gamma--ray bright AGN (iMOGABA) program, which is ideal for our test due to the fact that it observed up to 30 all--sky  gamma--ray bright radio-loud AGNs with fluxes ranging from $\sim0.2$ to $\sim20$~Jy at three frequencies related by an integer factor (21.5, 43 and 86~GHz) simultaneously\footnote{Most iMOGABA observations include 4 frequencies, except those in 2014B season (September, 2014 to January, 2015) 129 GHz was not included due to receiver problems.}. 

The observation was performed on 2014 September 1st with the KVN for a total time of 20 hours. All sources were observed in the same run in a snapshot mode alternating different scans of about 5 minutes each. Each source was observed for 2--8 scans. Cross-scan observations for pointing correction were conducted on every scan and sky-dipping curve measurements for sky opacity corrections were conducted every hour. All antennas recorded with a bandwidth of 64~MHz at 21.5 and 43~GHz and 128~MHz at 86~GHz, separated into different IF channels, with 16~MHz bandwidth each. Data were correlated with the DiFX software correlator at the Daejeon correlation center. Typical weather conditions were found during the observation, with averaged system temperatures $T_{sys}$ of the order of 200~K at 21.5~GHz, 200~K at 43~GHz, and 500~K at 86~GHz. For more information about iMOGABA observations see e.g. \citet{iMOGABA-I_paper} and \citet{Algaba15}.


The data analysis was performed using ParselTongue Python interface for the Astronomical Image Processing System (AIPS)~\citep[][]{parseltongue}. Amplitude calibration was performed following standard procedures with the task AIPS task APCAL. For phase calibrations, we first applied the earth orientation parameter (EOP) correction and parallactic angle correction. Then a prime calibration using 30 seconds data of 3C 454.3 (at UT 15:10:30) was performed for all IFs independently and the solutions were applied to all time ranges. The prime calibration lined up the phase at different IFs, and it also calibrated all the instrumental and atmospheric phase errors in the direction of the 3C 454.3 at this time. After the prime calibration, the data were split into single frequency sub-sets. 

To perform the FPT, we fringe--fitted the data for each source independently at 21.5~GHz ($\nu_{1}$) using a solution interval of 20 seconds with the AIPS task FRING. Point source models were used given that most sources show point-like or core-dominated structures under KVN resolution at this frequency~\citep{iMOGABA-I_paper}. The phase solutions obtained at 21.5~GHz were then scaled up by the corresponding frequency ratio and applied to the simultaneously measured data of the source itself at 43 GHz ($\nu_2$) and 86 GHz ($\nu_3$) to obtain the FPTed phases of $\phi^{\nu_2}_{A,FPT}$ and $\phi^{\nu_3}_{A,FPT}$, respectively, using a ParselTongue script which is a python implementation of  $use\_only\_mb.pl$ ~\citep{RD2014}. The 43 and 86 GHz FPTed residual phases are shown in Figure~1 and the left plots of Figure~\ref{fig:fpt2all}, respectively. 

Fringe-fitting was performed for the 43 GHz FPTed phase ($\phi^{\nu_2}_{A,PFT}$, as shown in Figure 1 using the whole 64 MHz bandwidth and a solution interval of 60 seconds\footnote{We can use a longer solution interval after the FPT due to the increase of the coherence time \citep[see e.g.][]{Algaba15}}. The obtained antenna-based phase solutions were calculated in ParselTongue based on the real and imaginary values in the resultant solution (SN) table as shown in the left plot of Figure~\ref{fig:snfpt}, with Yonsei as the reference antenna.
Then the phase values were scaled up by $R=2.5$ and corresponding real and imaginary values were calculated and written in a new SN table for the 86 GHz dataset using the {\it attach\_table} function in ParselTongue as shown in the right plot of Figure~\ref{fig:snfpt}. We note that, although the right plot in Figure~\ref{fig:snfpt} does not show any consideration of $2\pi$ unwrapping of the phase solutions, our analysis has considered such ambiguities using both script and manual check to avoid phase jumps due to the non-integer scaling factor.



\begin{figure*}
\label{fig:1}
\begin{center}
\includegraphics[angle=270, width=8cm]{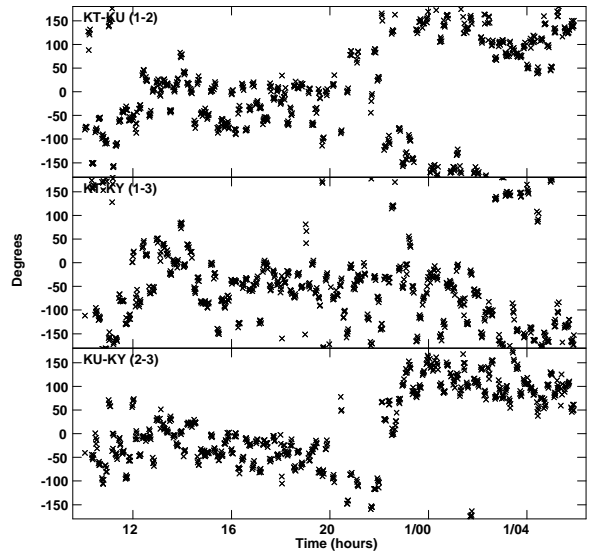}
\caption{Residual phases at 43~GHz ($\nu_{2}$) after FPT from 21.5~GHz ($\nu_{1}$) for the whole iMOGABA~16 observation, all 30 sources. This plot correspond to the reference dataset for FPT-square. From top to bottom, baselines are Tamna--Ulsan (KT--KU), Tamna--Yonsei (KT-KY) and Ulsan--Yonsei (KU-KY).}
\end{center}
\end{figure*}

After FPT--square, further extra steps were independently carried out to calibrate the remaining instrumental effects by either self-calibration or applying the calibrations from one or more sources using temporal interpolation.


\begin{figure*}
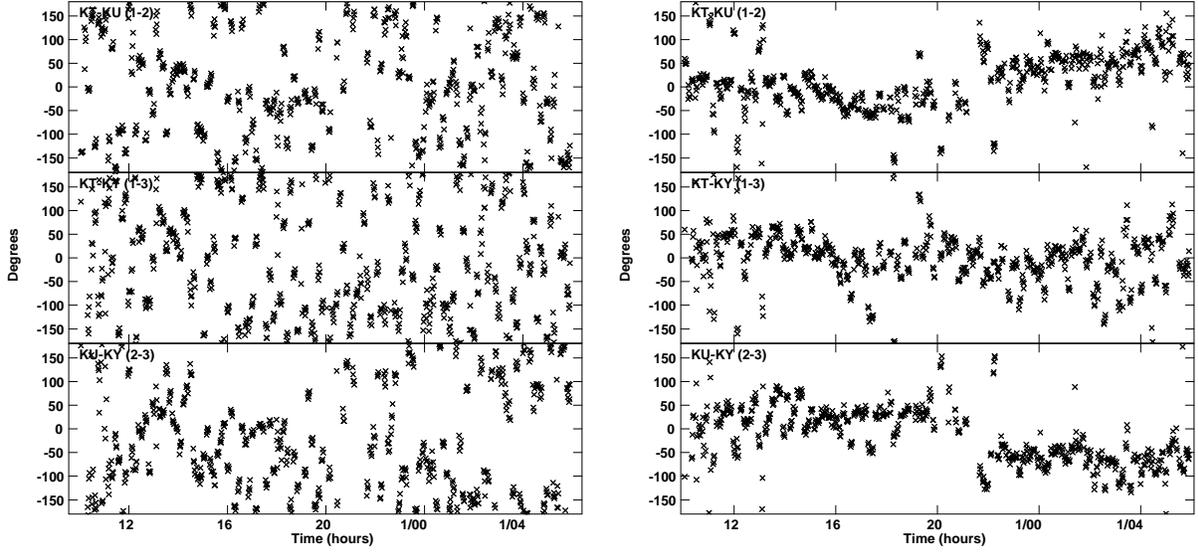

\includegraphics[angle=270, width=8cm]{fig3a.ps}
\includegraphics[angle=270, width=8cm]{fig3b.ps}
\caption{Residual phases at 86~GHz ($\nu_{3}$) for the whole iMOGABA~16 observation, all 30 sources. Left: after FPT from 21.5~GHz ($\nu_{1}$), i.e. before FPT--square; Right: After FPT--square. }
\label{fig:fpt2all}
\end{figure*}


\begin{figure*}
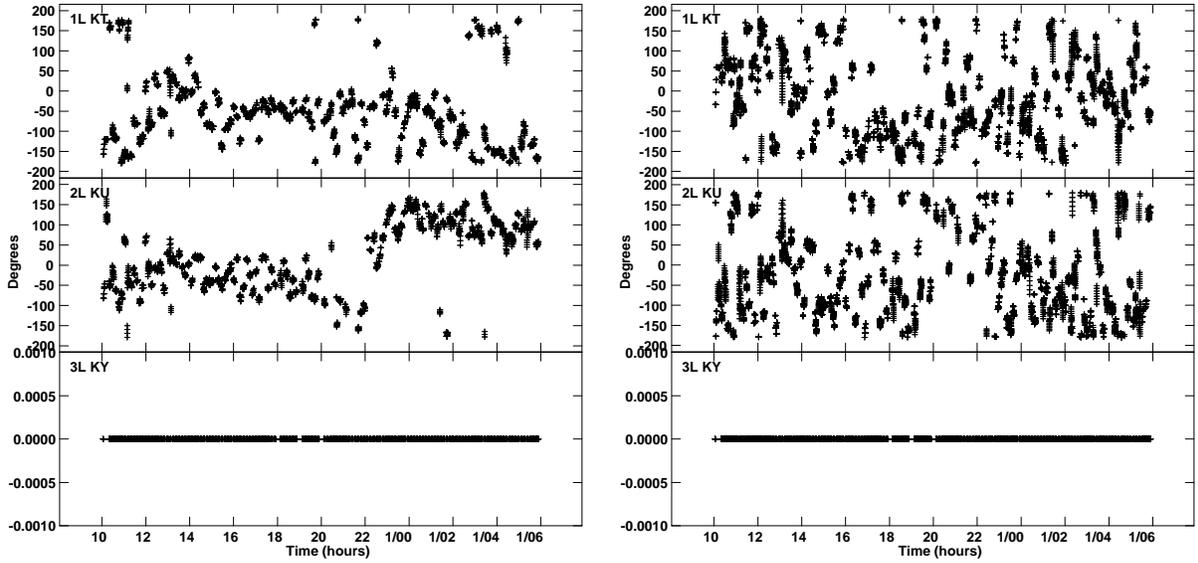

\includegraphics[angle=270, width=8cm]{sn8_q.ps}
\includegraphics[angle=270, width=8cm]{sn8_w.ps}
\caption{Phase solutions for FPT-square. Left: Solutions obtained by self-Fringe-fitting of the data at 43~GHz. KY is used as reference antenna. Right: Same as left, scaled by $R=2.5$. The plots to the right are the solutions for Figure~\ref{fig:fpt2all}, right}.
\label{fig:snfpt}
\end{figure*}

\section{Results}
\label{sec:results}

\subsection{Residual Phases}
Typical observations at high frequencies (such as at 86~GHz) have relatively short coherence times, of the order of just few seconds. It has been shown that FPT techniques can extend these values and it is expected that, for experiments such as iMOGABA, a longer coherence time, of the order of 20 minutes or more, can be obtained \cite[see e.g.][for observations up to 130 GHz]{RD2015, Algaba15}, resulting in residual phases with a small scatter within a typical scan length.

We note however that in snapshot mode observations, a significant number of sources with different lines of sight, is yet a limiting factor, and the residual phases still show a significant scatter when considered scans across the whole experiment (e.g. see Figure. \ref{fig:fpt2all}, left). This implies that, whereas the correction of tropospheric effects via FPT is very effective, there is certainly a residual offset arising from other components, i.e. ionosphere and instrumental.

As we correct for ionospheric effects via FPT--square, 
we obtain a much smaller scatter at 86 GHz even when we consider a large number of sources with a wide range of lines of sight (see Figure. \ref{fig:fpt2all}, right). It is clear from comparison of both plots that the offset of the residual phases between different scans has significantly decreased and phases follow a more clear pattern along the experiment. This shows that ionospheric effects are not always negligible at these frequencies.


In Figure. \ref{fig:fpt2all} right, it seems that some of the residual phases have a non--negligible offset for some sources. In some cases, such as at the beginning of the observations in Ulsan--Yonsei and near the end in Tamna--Yonsei baselines, this offset leads to a possible double layer. This can be associated with two different groups of sources observed at similar time ranges with a very large difference ($>40\degr$) in declination. Two sources observed in this time range, Sgr A* and 1921-293, have even larger phase offset due to the same reason. In the case of  3C~84 which was observed for several scans between UT 15:00 and 22:00, the additional phase offset arises due to a mismatch of the peak position at different frequencies. The brightest component we see at 86~GHz and lower frequencies does not correspond to the same feature due to free-free absorption \citep[][]{WalkerApJ...530...233}; See also~\citet{iMOGABA-I_paper}. 

Nonetheless, the general trend is very clear: even for the sources with large separation, residual phases vary smoothly over several hours, indicating a very long stability, even at such high frequency. We discuss this in more detail below.

\subsection{Increase of the Coherence Time}
\label{sec:coherencetime}
As we have previously mentioned, a direct consequence of the compensation of the residual phases via FPT and FPT--square is the increase of the coherence time and, therefore, the sensitivity. From visual inspection of Figure \ref{fig:fpt2all}, it seems clear that the coherence time has increased to more than a few tens of minutes.

\begin{figure}
\includegraphics[angle=-90, origin=c , width=6.9cm]{fig4a.ps}
\includegraphics[height=7.3cm]{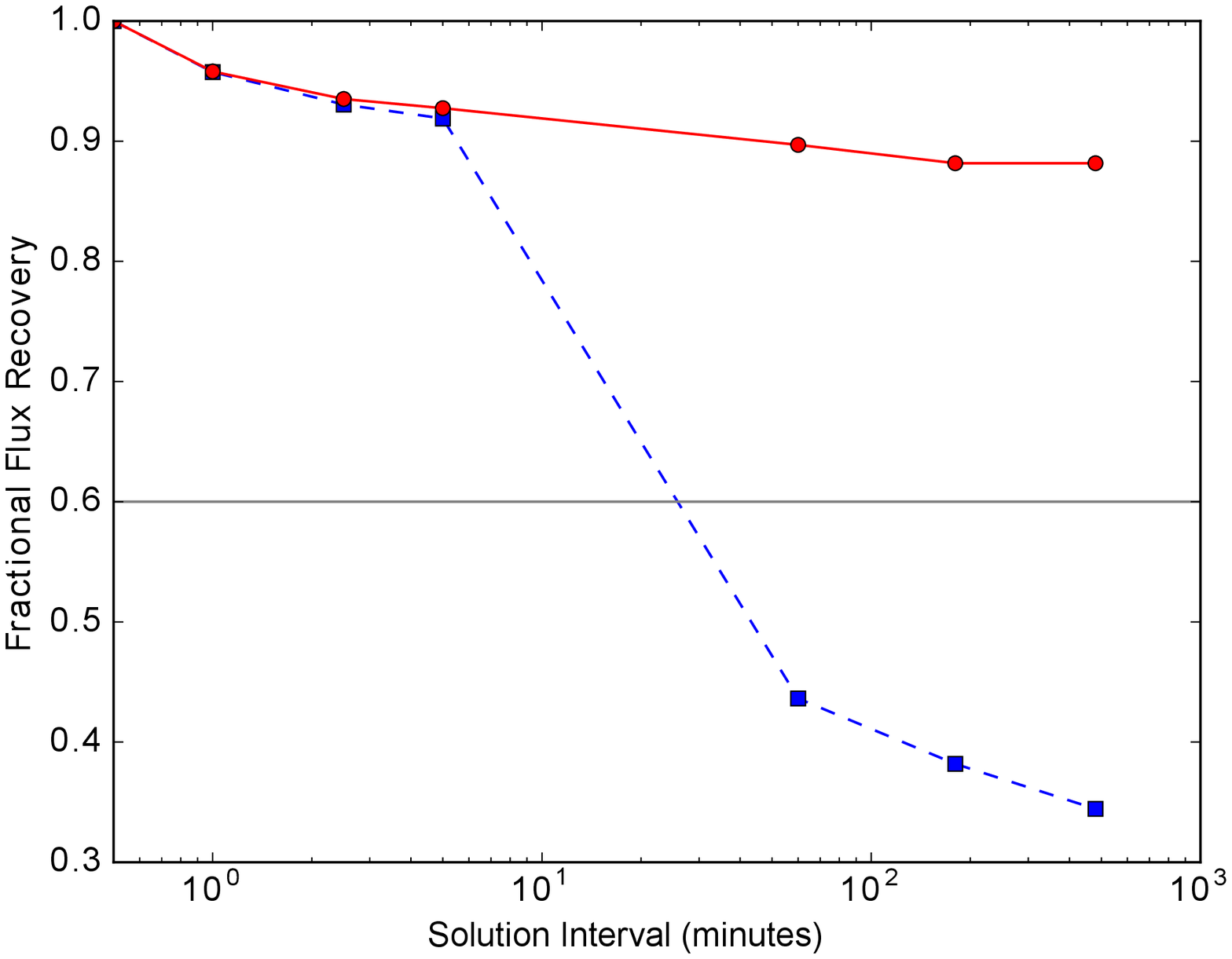}
\caption{Left: Residual phases of OJ~287 at 86~GHz after FPT-square. Right:} Peak flux recovery for OJ~287 at 86~GHz. Blue dotted and red straight lines indicate the recovery for FPT and FPT--square respectively. Dots indicate actual measurements, whereas the lines are for eye guidance. Grey horizontal line show a recovery of 0.6, consistent with a 1 rad phase rotation.
\label{fig:coher_time}
\end{figure}

In order to quantify the increase of the coherence time with FPT--square, we follow a similar procedure as the one in \citet{RD2015}. We perform the analysis of the coherence times achieved with FPT--square at 86~GHz. A summary of the method is as follows. We used the AIPS task CALIB on the FPT--square data sets of OJ~287 which provides the maximum source time interval, up to 8 hours, in our observation(see Figure~\ref{fig:coher_time}, left). We then self--calibrated the data with different solutions intervals, produced maps by Fourier--transforming the visibilities, and calculated the fractional recovered peak flux, which is the peak flux at a given time interval weighted by the peak flux at the shortest time interval; i.e. 30 seconds. 

In the right panel of Figure \ref{fig:coher_time} we plot the fractional peak recovered flux for the different time intervals. As expected, it slightly decreases as the cumulative time increases. If we define the coherence time as the solution time for which the peak flux recovery is 60\% which is equivalent to the rms residual phase being 1~rad \cite[see e.g.][]{RD2015}, it is clear that the coherence time after FPT--square surpasses 8~hours, with a fractional recovery above 88\% even at such long time interval.

For comparison purposes, we also explore the coherence time at 86~GHz achieved with normal FPT calibrated with 21.5~GHz. The fractional peak recovered fluxes for FPT are also shown in Figure \ref{fig:coher_time}. In this case it seems that the coherence time is longer than the scan length (5~minutes), in agreement with the discussion in \cite{Algaba15}, but shorter than the shortest scan gap found for this experiment (1 hour). This clearly shows the improvement of the coherence time with FPT--square over typical FPT by at least an order of magnitude.

\subsection{Using a Second Source for Eliminating Residual Instrumental Time Variations}

From Equation \ref{eq:FPTsqr}, it is clear that, after FPT--square, instrumental effects are still present. These residual phases are zeroed at the time of prime calibration but still have slow time variations as can be seen from Figure~\ref{fig:fpt2all}, right. Nevertheless, the significant improvement in the phase residuals both in terms of scatter and stability allows us to consider additional operations such as using the calibration of another source for compensating these variations. Based on the encouraging results seen in Figure \ref{fig:fpt2all}, we would be able to do this for source pairs with large angular separations. This will further extend the coherence time. 
We note that, calibrating with another source usually implies astrometry, and great efforts have been made in the past, with conventional phase-referencing\citep[e.g.][]{BeasleyConway95} and SFPR~\citep{RD2011}. However, our method does not achieve astrometry (see Equation~\ref{eq:FPTsqr}). Nevertheless, weak source detection at high frequencies could benefit from this method.


As a general rule, we select a source pair with the following criteria: i) relatively good number of scans in this observations, ii) the calibrator should be a bright source with compact structure, and iii) the target should ideally have some well-known structure that we can still retrieve by the phase compensation for a consistency check. We consider various cases in the following subsections.

\subsubsection{$10\degr$ separation: 3C~279 and 3C~273}

As our first analysis, we study the feasibility of phase compensation after  FPT--square between two sources separated by $10\degr$. Even at lower frequencies and under very nice observing conditions, a conventional phase-referencing for a separation angle larger than a few degrees is not in general an easy process because of the line of sight dependence of the atmospheric effects.

3C~279 is a highly polarized quasar at a redshift of $z=0.536$ \citep[][]{3c279z} with flat spectrum and a flux density of up to $\sim10$~Jy at 86~GHz. Despite showing a clear jet structure on the south--west direction \citep[e.g.][]{MOJAVE}, it is not resolved with KVN baselines, making its structure relatively compact for this observation and an ideal source as a calibrator \citep[Figure.~\ref{fig:mapspair1}; see also][]{iMOGABA-I_paper}.

On the other hand, 3C~273, at an angular separation from 3C~279 of $10.4\degr$, is a flat spectrum quasar at $z=0.1583$ \citep{Strauss92} with flux density of about 3~Jy at 86~GHz whose structure is partially resolved by KVN observations and can be described as a core plus a jet component located at around 2~mas to the southwest~\citep{iMOGABA-I_paper}. 

\begin{figure*}
\includegraphics[angle=0, width=8cm]{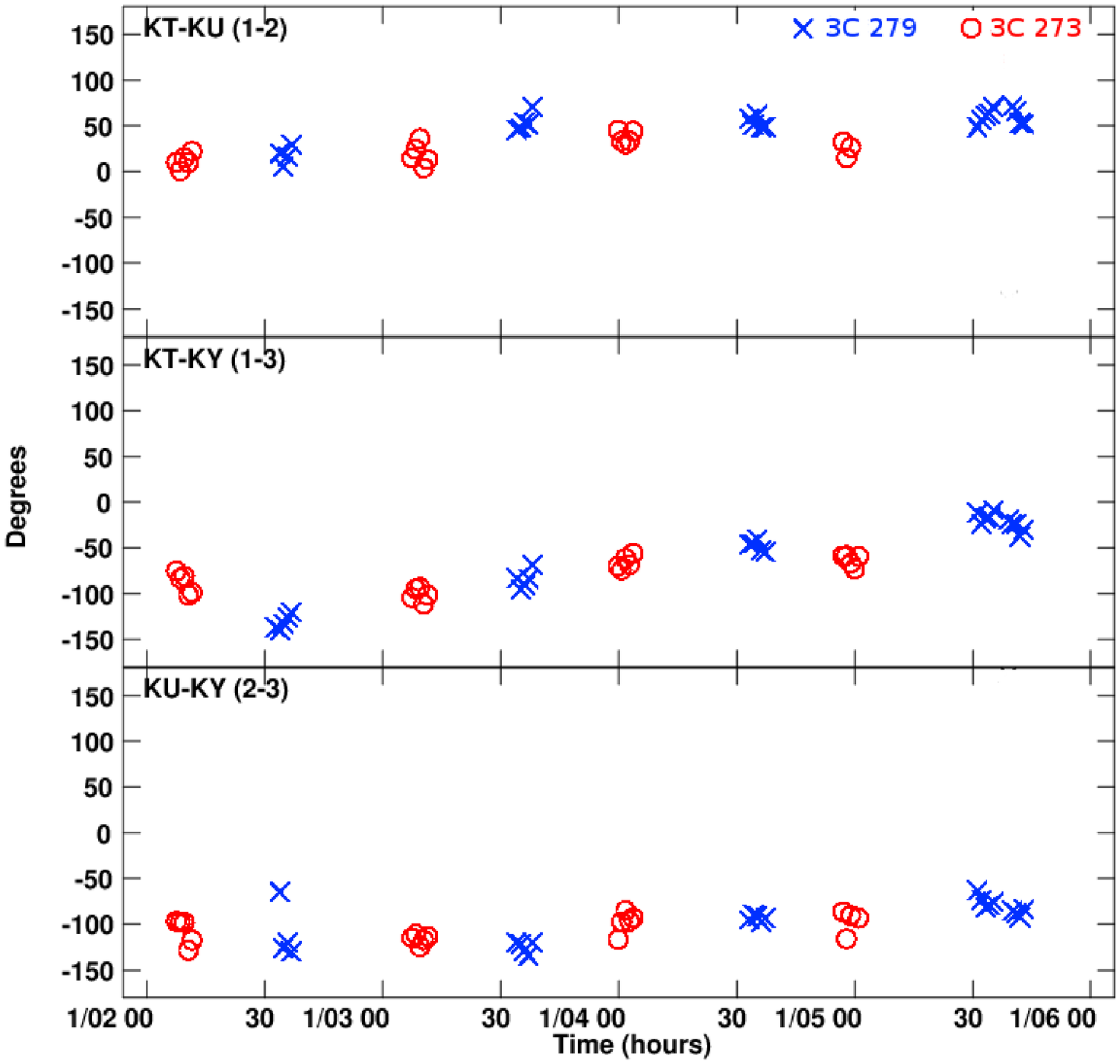}
\includegraphics[angle=0, width=8cm]{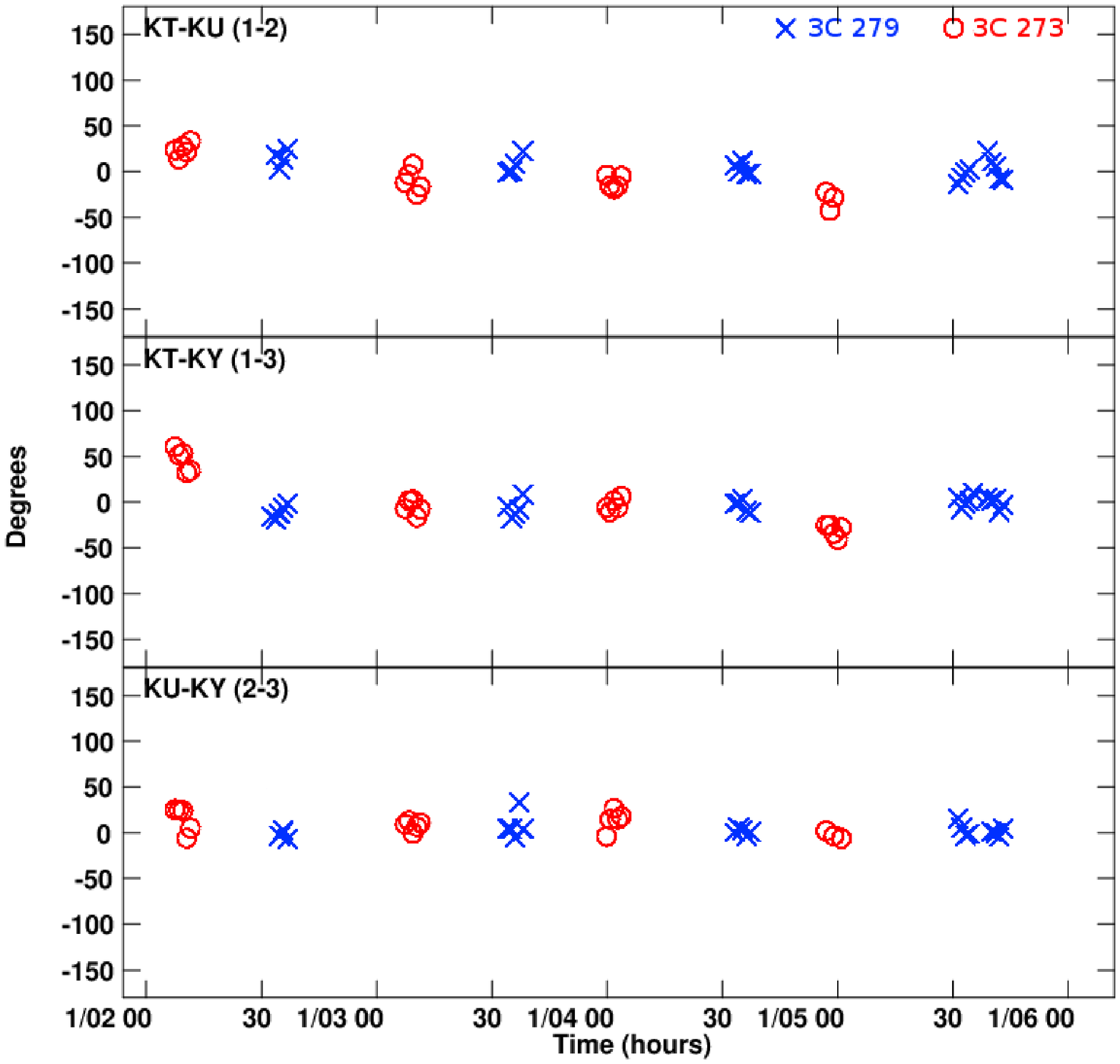}
\caption{Residual phases for the source pair 3C~279 (blue crosses) and 3C~273 (red circles) at 86~GHz. Left: after FPT--square; right: after fringe--fitting on 3C~279 and applying the solutions to both sources (using temporal linear interpolation for 3C~273).}
\label{fig:fpt2pair1}
\end{figure*}

\begin{figure*}
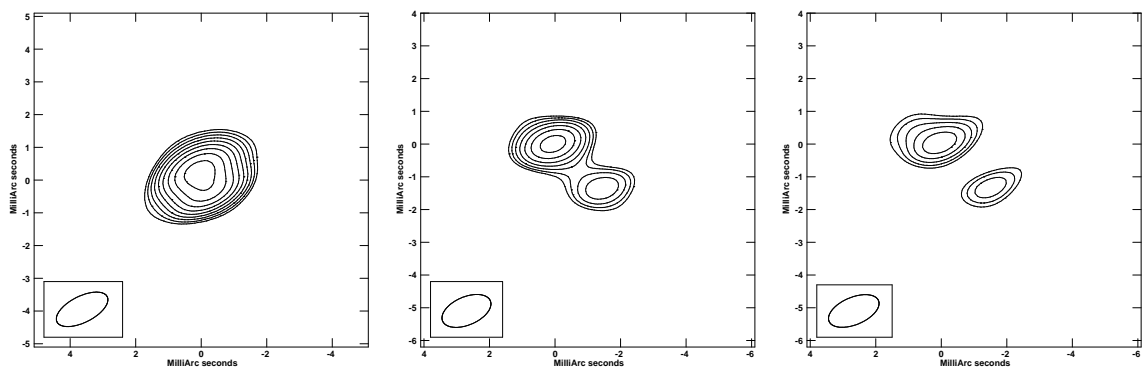
   
\includegraphics[angle=270, width=5cm]{3c279wselfc.ps}
\includegraphics[angle=270, width=5cm]{3c273wselfc.ps}
\includegraphics[angle=270, width=5cm]{3c273wsf2pr.ps}
\caption{Demonstration of phase compensation after FPT--square at 86~GHz with calibrator at $10\degr$ separation. Left: hybrid map of the calibrator 3C~279 after FPT--square. Middle: hybrid map of the target 3C~273 after FPT--square. Right: CLEANed map of the target 3C~273 after FPT--square and applying the calibration of 3C~279. The rms noise values in the maps are 94, 66, and 104 mJy/beam, respectively. The contours in each panel start from 3 times of the rms level and increase by a factor of $\sqrt{2}$. The ellipse on the bottom left of each panel shows the natural weighted beam size.}
\label{fig:mapspair1}
\end{figure*}

Figure \ref{fig:fpt2pair1} (left) is an excerpt of Figure \ref{fig:fpt2all} (right) where we only focus on the scans corresponding to the source pair 3C~279 and 3C~273. The sources were observed for 5 and 4 scans respectively, during a period of 3 hours. 
The gap between the scans of 3C~273 and 3C~279 is about 30~minutes (i.e. a duty cycle of 1 hour). As expected from the results shown for all sources in Figure \ref{fig:fpt2all}, it is clearly seen that, after the FPT--square, the phase residuals of the two sources follow a similar trend, with the residual phases on one closely following these of the other source. It is thus clear that, although the observation is a-priori not optimized for source switching, after FPT--square it is straightforward to conduct the calibration between them by temporal interpolation of phase solutions. 

When we calibrate 3C~279 and apply its phase solutions to 3C~273, all residual phases are well calibrated (Figure \ref{fig:fpt2pair1},right), the phases of 3C~273 are mostly within $\pm30\degr$ degrees around zero for all baselines. A good way to check for the goodness of this procedure is to produce a map, using just a
Fourier transform of the visibility data, and compare it with the self--calibrated one (see Figure~\ref{fig:mapspair1}). The jet structure is successfully recovered, although no additional self--calibration was done on the data for the mapping. The peak flux recovery is around 88\%, indicating a very good phase--compensation even this observation is not scheduled for such kind of analysis.

\subsubsection{$20\degr$ separation: OJ~287 and 4C~39.25}

As our next step, we explore the possibility of residual phase compensation between two sources with an angular separation of up to $20\degr$. We emphasize here that, this would be the largest angular separation that has been tried so far. We choose OJ~287 as the calibrator and 4C~39.25 as the target, given their structure and flux density.

OJ~287 is a  relatively compact BL Lac object at redshift $z=0.306$ \citep{1989A&AS...80..103S} and have a flux density above 5~Jy at 86~GHz. Its structure appears as point--like under the view of the KVN array, which allows us to use this source as a good calibrator.  

4C~39.25, at an angular separation of $20.2\degr$ from OJ~287, classified as a flat spectrum radio quasar at a redshift of $z=0.695$ \citep{2004AJ....128..502A}, and with a flux density of about $\sim2.5$~Jy at 86~GHz, has a well--known peculiar core--jet structure, where its brightest component even at cm wavelengths corresponds to the optically--thin jet, whereas the additional dimmer structure on the west corresponds to its radio--core \citep[e.g.][]{1997A&A...327..513A}. 

\begin{figure*}
\includegraphics[angle=0, width=8cm]{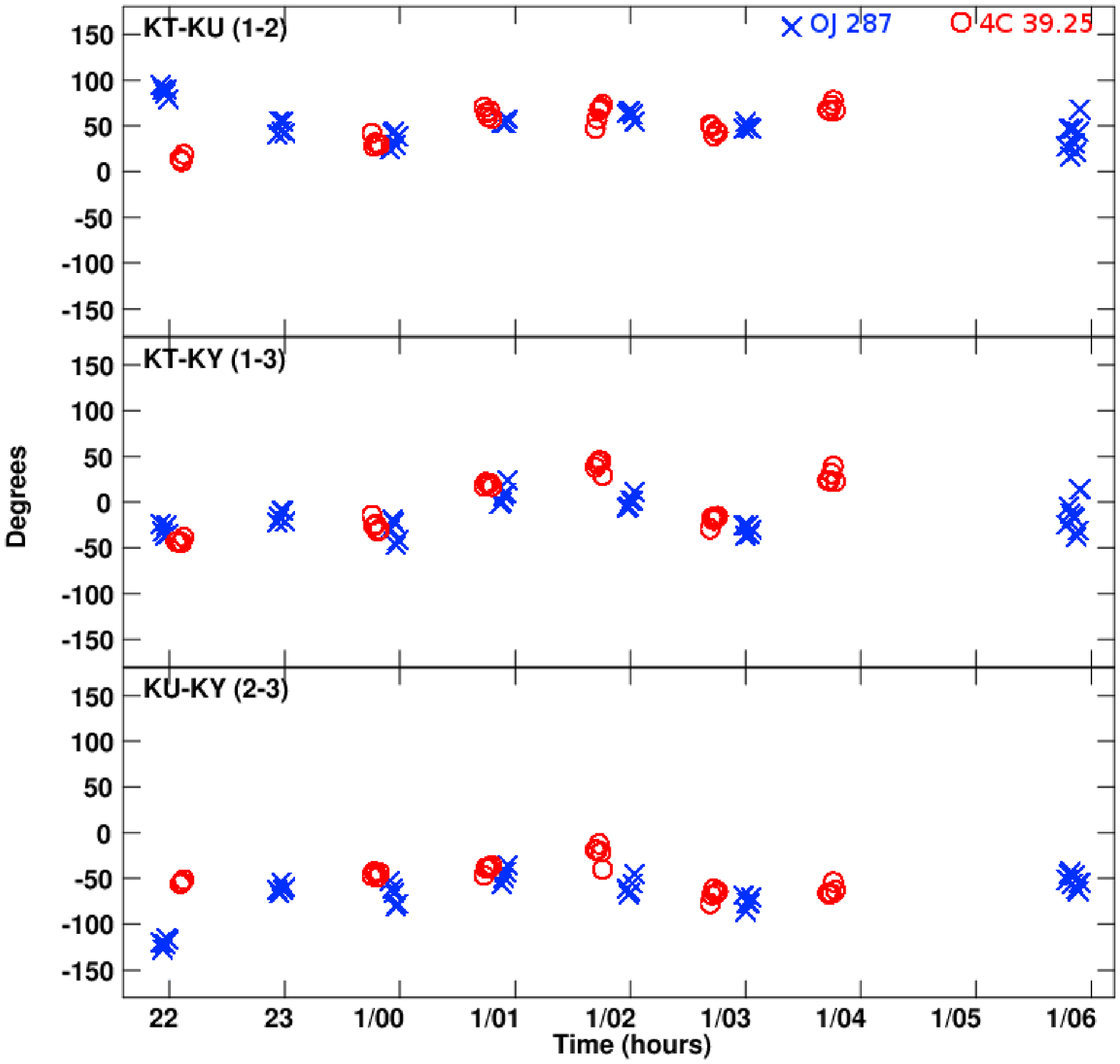}
\includegraphics[angle=0, width=8cm]{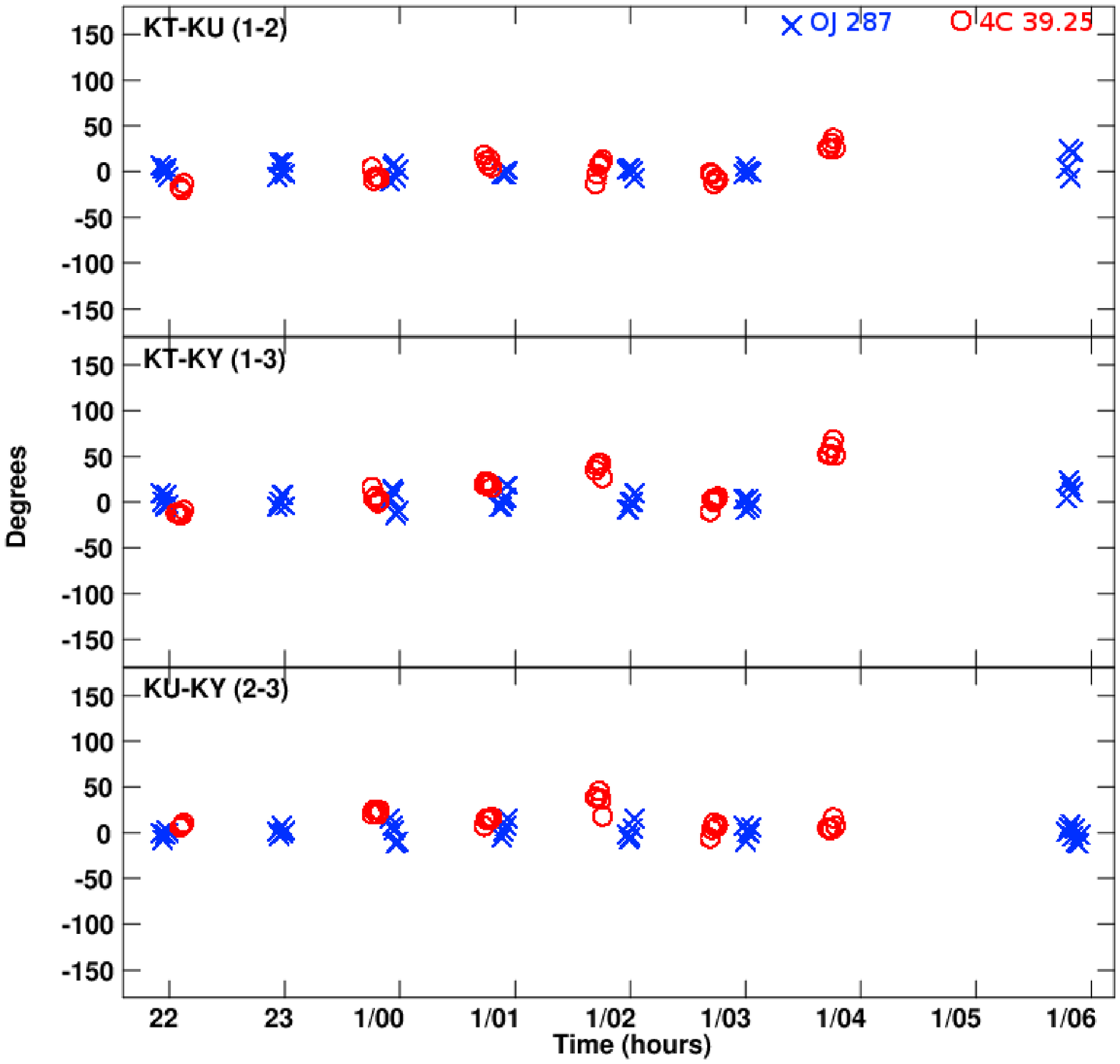}
\caption{Residual phases for the source pair OJ~287 (blue crosses) and 4C~39.25 (red circles) at 86~GHz. Left: after FPT--square; right: after fringe--fitting on OJ~287 and applying the solutions to both sources (using temporal linear interpolation for 4C~39.25).}
\label{fig:fpt2pair2}
\end{figure*}

In Figure \ref{fig:fpt2pair2} (left) we show the scans where these two sources were observed. OJ~287 was observed for 7 scans within 8 hours, whereas 4C~39.25 was observed for 6 scans during 6 hours. The gap between the scans of these two sources range from few minutes to two hours, making such observation particularly inadequate for temporal interpolation. Nonetheless, after the FPT--square, the phase residuals of the two sources follow a well--behaved smooth variation with a very similar trend. Only for the first scan on Ulsan baselines there are some offsets on OJ~287 residual phases. 
This offset is due to a reset of the antenna controlling system after this scan, leading to different instrumental phase values. This was caused by the unusual antenna operation. FPT-square process does not intrinsically induce such abrupt phase change.

\begin{figure*}
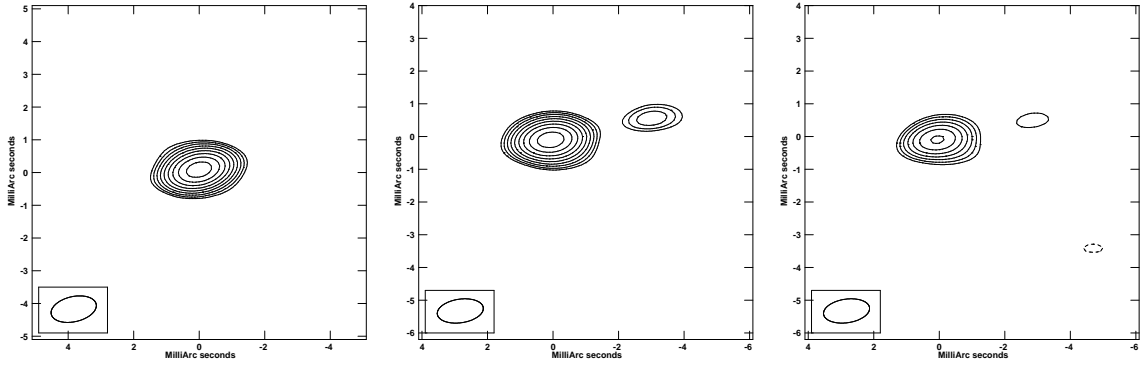
  
\includegraphics[angle=270, width=5cm]{oj287wselfc.ps}
\includegraphics[angle=270, width=5cm]{4c39.25selfc.ps}
\includegraphics[angle=270, width=5cm]{4c39.25sf2pr.ps}
\caption{Demonstration of phase compensation after FPT--square at 86~GHz with calibrators at $20\degr$ separation. Left: hybrid map of the calibrator OJ~287 after FPT--square. Middle: hybrid map of the target 4C~39.25 after FPT--square. Right: CLEANed map of the target 4C~39.25 after FPT--square and applying the phase calibration of OJ~287. The rms noise values in the maps are 46, 32, and 74 mJy/beam, respectively. The contours in each panel start from 3 times of the rms level and increase by a factor of $\sqrt{2}$. The ellipse on the bottom left of each panel shows the natural weighted beam size.} 
\label{fig:mapspair2}
\end{figure*}

We calibrated the phases of OJ~287 and applied the solutions (excluding the first scan) to 4C~39.25. This leads to quite well aligned phases around $0\degr$, with an offset within $\pm50\degr$, similar to the one obtained for the previous source pair. In Figure \ref{fig:mapspair2} (middle and right) we compare the cleaned map with the one produced by self--calibration. The main structure is properly retrieved with over 3-$\sigma$ confidence level and the flux recovery is around 97\% indicating that, even for such large angular separation, phase compensation is feasible after FPT--square.

\subsection{All-sky Phase Compensation using more than one Calibrator}

Given the results in the previous section, it is clear that, in the most general case, we can apply phase compensation between source pairs with even larger separations. 
We note however that, with the schedule of the experiments that we discuss here, it is not easy to find source pairs with larger separation but with a good scan overlap between sources (i.e. scans between different sources separated by less than a few hours). Instead, we follow an alternative approach, which is to select a few limited number of bright and compact sources and use them to calibrate the rest of the targets in the experiment by linear temporal interpolation, with RA spanning from 0 to 24 hours.

We chose the calibrators as follows: In order to cover all sky, we split it in various regions in RA, each of them containing a calibrator as follows: 3C~111 for RA 0--6~hours; 4C~39.25 for RA 6--12~hours; 3C~345 for RA  12--18~hours and 3C~454.3 for RA 18--24~hours. We used these sources to calibrate the rest of the sample by temporal interpolation. 
For the few sources with lower declination (DEC $<5\degr$; 0420--014, 0727--115, 3C~273, 3C~279, and NRAO~530),  
we used 0420--014 and 3C~279 as calibrators. 
We note that, with this selection, in some extreme cases (e.g. 1308+216, which is located between 4C39.25 and 3C 345) the separation between the source and the nearest calibrator is very large, up to $45\degr$, more than twice of the ones discussed above.

\begin{figure}
\begin{center}
\includegraphics[angle=270, width=8cm]{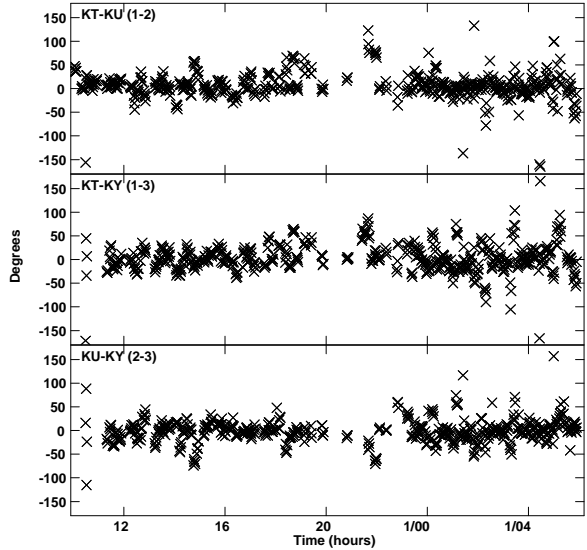}
\caption{Residual phases at 86~GHz after FPT--square and phase compensation with multiple calibrators for the whole observation.}
\label{fig:sf2prall}
\end{center}
\end{figure}

In Figure \ref{fig:sf2prall} we plot the residual phases after the compensation for the whole observation. Here we excluded Sgr A* and 1921--293 because of their very low elevation ($<$ 30 degrees), giving rise to very low signal to noise ratios during the FPT and FPT--square and large instrumental errors; and 3C~84 because the model solutions were not appropriate for this source, as the brightest component we see at different frequencies does not correspond to the same feature.

Residual phases are very well aligned thought the whole experiment for all baselines, with most of them within $\pm60\degr$ ($\sim$1~rad), indicating the good phase coherence achieved in FPT--squared data.

\section{Discussion}
\subsection{Characteristics of Residual Phases---The Power of Simultaneous Multi-frequency Observing}
The use of simultaneous multi--frequency observations opens a new path in the calibration of high frequency data and the compensation of the phase coherence deteriorated by propagation through the atmosphere. By means of FPT, it is possible to obtain phase solutions at a lower frequency and apply these to a higher frequency, effectively correcting for the tropospheric effects, leading to an increase of the coherence time and thus the sensitivity. As a further step, FPT--square makes use of two low frequencies to calibrate the residual phases of a higher frequency by correcting in practise for both tropospheric and ionospheric effects. This further increases the coherence of the phases to several hours thus boosting the sensitivity and allowing the use of additional techniques such as compensation using other sources without their typical limitations on separation and switching cycle.

With the use of FPT--square, we have been able to perform phase compensation between sources with the largest separation so far although the schedule of our observations was not at all optimized for source-switching.

As can be seen from Figures \ref{fig:fpt2pair1} and \ref{fig:fpt2pair2}, after FPT--square and source pair phase compensation, the phases are aligned with similar dispersion ($\sim50\degr$) for both cases of source pairs with  10 and 20 degrees angular separation respectively. It is possible that, for the case of $20\degr$ angular separation, the dispersion is slightly smaller. As we map the target sources, we observe an 88\% flux recovery for the $10\degr$ pair, whereas we find a 97\% recovery for the $20\degr$ pair.

This suggests that angular separation may not be the most important factor and there may be other factors to consider when conducting phase compensation between sources. For example, it is clear that the switching cycle has produced a better sampling interval between target and calibrator for the pair OJ~287 and 4C~39.25. The temporal variation of the instrumental effects would be the reason for this. Thus, scheduling, time gaps between scans, total number of scans and visibilities and such factors can also play a role in a reliable structure identification and flux recovery. This indicates that we can indeed find source pairs with even larger angular separation and perform a reliable phase compensation as long as we take these additional factors into account.


We note that a stable array performance would be very important for a successful FPT--square experiment because its effects are also transferred during FPT-square, i.e. the instrumental term in Equation~\ref{eq:FPTsqr} is different as that in Equation~\ref{eq:FPT13}. If it becomes smaller, this could also contribute to the increase of coherence. However, if the instrumental effects dominate over the ionospheric ones in the FPTed phases, even the later ones are calibrated via FPT-square, the phase coherence will not increase. The implementation of multi-frequency P-cal system will help to ensure the instrumental stability (See Section~\ref{sec:mfpcal}).

\subsection{Extension of Coherence Time Achieved with FPT--square}

In Section \ref{sec:coherencetime} (see also Figure \ref{fig:coher_time}), we have demonstrated that we can obtain a coherence time longer than 8~hours for a 86~GHz VLBI experiment via FPT--square. In the following we will discuss the relative improvement and implications when compared with previous studies.

For normal single--frequency observations and standard data analysis, typical coherence time scales are smaller than these found here by more than an order of magnitude. Characteristic coherence times for VLBA are of the order of 40 minutes at frequencies around 2~GHz \citep[see e.g.][]{TMS01} and fall below few tens of seconds at frequencies around 86~GHz.

As we have discussed above, short intervals between target and calibrator scans are desired for successful observations, such large coherence time implies that we can perform comparatively much longer switching cycles between sources, up to few hours, without significant loss of coherence. By comparison, typical switching times in conventional phased--referencing observations would be of the order of few minutes or less, while with SFPR, the switching cycle can be extended up to several minutes even at 130~GHz \citep[e.g.][]{RD2015}.

\subsection{Broad Scope of Applicability}

In general, residual phases of high frequency observations are hard to calibrate, with conventional techniques such as fast source switching not feasible, due to the short coherence times. Additional techniques or methodologies, such as FPT make it possible. Among the methods based on FPT, SFPR requires a nearby calibrator within 10 degrees. This implies that the residual phase will contain ionospheric interpolation errors arising from different lines of sight. The instrumental phase errors are also calibrated in SFPR \citep{RD2011}. On the contrary, both MFPR and FPT-squared need additional steps to remove the instrumental phase effects~\citep[e.g.][]{Dodson17}, but since the ionospheric residuals are corrected for along the same line of sight of the target, the calibrator can be further away, up to few tens of degrees, and the temporal gap between observations of target and calibrator can be larger. In this sense, both MFPR and FPT-squared methods are suitable for the cases that a nearby phase calibrator is not found. MFPR can mitigate the ionospheric TEC error to $\sim$0.1 TEC Unit level by using a set of low frequency observations~\citep{Dodson17}. Currently no VLBI facility can work at all these frequencies simultaneously, which means fast switching and temporal interpolation are required in MFPR. FPT-square uses two lower frequencies for atmospheric calibrations and the reference frequencies can be higher compared with those in ionospheric execution blocks in MFPR and no interpolations are necessary as all frequencies are observed simultaneously. On the other hand, both SFPR and MFPR can achieve astrometry while in FPT-square the astrometry is contaminated during the multi-step phase transfer only a weighted combination of core-shifts are left (Eq.~\ref{eq:FPTsqr}). Furthermore, all methods subject to multi--step calibrations lead to an accumulation of thermal noise errors.

We note that there is no need of a particularly restrictive schedule guideline for FPT--square. For example, there is no need to search for nearby calibrators, as long as the sources are detected with a decent signal to noise ratio at the lower frequencies ($\nu_{1}$ and $\nu_{2}$). Furthermore, given the fact that each source's phase is calibrated with its own counterpart at lower frequencies, there is no need for caution on frequency or source switching cycles, thus avoiding the limitations imposed by them. In a similar way there is neither a particular restriction in terms of frequency. In this work we have used a particular set of frequencies to perform the FPT--square to up to 86~GHz but, as long as care is taken for the possible phase jumps due to the $2\pi$ ambiguity, this strategy can be used for other frequency setups. The results presented here indicate that the FPT--square technique would continue to work at even higher frequencies.

Simultaneous multi--frequency capabilities are becoming available in a significant number of facilities. Originally designed for the KVN, the feasibility and power of this multi--frequency band receiver system has been extensively proven and is now being installed and experimented in other facilities. Telescopes like Yebes and the VERA antennas are among the instruments that can perform simultaneous observations~\citep{2015JKAS...48..277J}. Future development in receivers is also pointing in the direction of compact multi--band receivers (Han et al. in prep). Alternatively, in the case of facilities consisting on connected antennas, such as ATCA, each frequency can be simultaneously observed by splitting the system into various sub--arrays. Note that if a facility does not have simultaneous multi--frequency capabilities, it is still possible to perform FPT--square with fast frequency switching, although time interpolation will be needed, leading to some additional random phase errors.

As discussed above, it becomes clear that one of the strengths and uniqueness of FPT--square is its suitability in the context of high-frequency all-sky surveys. The use of this technique will allow to image sources that have been yet undetected at 86~GHz and/or higher frequencies. For example, the Multi-frequency AGN Survey with KVN (MASK, Jung et al. in prep) can take advantage of this technique to further improve the detection rate at such frequencies.

\subsection{Towards Imaging with FPT--square by Implementation of P-cal System} 
\label{sec:mfpcal}

In section \ref{sec:results} we used additional sources to calibrate the residual instrumental variations. As we have already mentioned above, another possibility is to use a pulse--cal system (P--cal). Injection of a pulse calibration near the feeds every few micro--seconds will help to calibrate the bandpass phase characteristics and its time variation~\citep[e.g.][]{Thompson91}. P--Cal systems have been already implemented in facilities like the VLBA. Once a multi--frequency P--Cal system is implemented in the KVN, we will be able to correct instrumental phase variations such as LO drift and cable delays between bands, thus an improved calibration of the residual phases is expected (Je et al, in prep.). This implies that the full calibration of VLBI phase errors will be achievable with FPT-square by adding P--cal, and therefore imaging the source structure at the target frequency ($v_3$) after FPT--square will be obtained. Furthermore, we would like to note that, along with the source structure, a measure of the weighted combination of the core-shifts \citep[or ``centroid'' shift due to structure blending effects, see e.g.][]{RD2014} and higher accuracy of the position measurements of the centroid can be obtained.

\section{Conclusions}

We have demonstrated that, with the KVN simultaneous multi--frequency receiving system, we are able to further extend the FPT technique. By using a set of two lower frequencies to calibrate a higher one, we are able to correct not only the tropospheric but also the ionospheric effects. Instrumental effects are effectively the only phase residuals left in the data. At 86~GHz the coherence time was extended from about 20~minutes after FPT using 21.5~GHz data to more than 8~hours after FPT--square. This clearly enables us to improve the sensitivity in the detection of weak sources. 

The improvement of the phase coherence also allows us to carry out source--pair phase--compensation without tight constraints on switching cycle or source separation. By using one set of iMOGABA observations with more than 30 all--sky sources with snapshot mode, which schedule is not optimized for source switching, we have successfully detected targets separated up to $20\degr$ from the calibrator and our results suggest that the separation limit can be even larger than this. Additionally, we have also been able to use several calibrators and carry out phase compensation to all--sky sources. In the future, instrumental phase variations can be calibrated with the multi-frequency P-cal system.

One of the unique aspects of this calibration method is its suitability for high frequency all-sky surveys including weak sources. Any data obtained by simultaneous or fast--switching multi--frequency observations with good instrumental stability can be used for FPT--square. In this paper we have analyzed an iMOGABA epoch, but with the increase of facilities with simultaneous multi--frequency capabilities, similar analysis can be performed to other kind of observations, such as the MASK project (Jung et al. in prep).


\acknowledgments

The KVN is a facility operated by the Korea Astronomy and Space Science Institute. G. Zhao and T. Jung are supported by Korea Research Fellowship Program through the National Research Foundation of Korea (NRF) funded by the Ministry of Science and ICT (NRF-2015H1D3A1066561). S. S. Lee and S. Kang were supported by the National Research Foundation of Korea (NRF) grant funded by the Korea government (MSIP) (No. NRF-2016R1C1B2006697). D.-W. Kim and S. Trippe acknowledge support from the National Research Foundation of Korea (NRF) via grant NRF-2015R1D1A1A01056807. J. Park acknowledges support from the NRF via grant 2014H1A2A1018695.

\end{document}